\documentclass[sigconf]{acmart}

\usepackage{subfig} 
\usepackage{multirow}   
\usepackage{arydshln} 
\usepackage{enumitem}

\usepackage[show]{notes}
\newcommand{\rp}{\texttt{r/place}}
\newcommand{\ignore}[1]{}
\newcommand{\com}[1]{}

\AtBeginDocument{%
  \providecommand\BibTeX{{%
    \normalfont B\kern-0.5em{\scshape i\kern-0.25em b}\kern-0.8em\TeX}}}

\setcopyright{acmcopyright}
\copyrightyear{2021}
\acmYear{2021}
\acmDOI{10.1145/1122445.1122456}

\ignore{
    \acmConference[Woodstock '18]{Woodstock '18: ACM Symposium on Neural
      Gaze Detection}{June 03--05, 2018}{Woodstock, NY}
    \acmBooktitle{Woodstock '18: ACM Symposium on Neural Gaze Detection,
      June 03--05, 2018, Woodstock, NY}
    \acmPrice{15.00}
    \acmISBN{978-1-4503-XXXX-X/18/06}
}

\acmConference[WWW '22]{WWW '22: ACM Symposium on the Web}{April 25--29, 2022}{Lyon, France}

\copyrightyear{2022}
\acmYear{2022}
\setcopyright{acmlicensed}\acmConference[WWW '22]{Proceedings of the ACM Web Conference 2022}{April 25--29, 2022}{Virtual Event, Lyon, France}
\acmBooktitle{Proceedings of the ACM Web Conference 2022 (WWW '22), April 25--29, 2022, Virtual Event, Lyon, France}
\acmPrice{15.00}
\acmDOI{10.1145/3485447.3512238}
\acmISBN{978-1-4503-9096-5/22/04}

\begin{document}

\title{This Must Be the Place: Predicting Engagement of Online Communities in a Large-scale Distributed Campaign}

    \author{Abraham Israeli, Alexander Kremiansky, Oren Tsur}
    \email{{isabrah,kremians}@post.bgu.ac.il, orentsur@bgu.ac.il}
    \affiliation{
      \institution{Department of Software and Information System Engineering\\ Ben-Gurion University of the Negev}
      \country{Israel}
    }
    

\begin{abstract}
    Understanding collective decision making at a large-scale, and elucidating how community organization and community dynamics shape collective behavior are at the heart of social science research. In this work we study the behavior of thousands of communities with millions of active members. We define a novel task: predicting which community will undertake an unexpected, large-scale, distributed campaign.
    To this end, we develop a hybrid model, combining textual cues, community meta-data, and structural properties. We show how this multi-faceted model can accurately predict large-scale collective decision-making in a distributed environment. We demonstrate the applicability of our model through Reddit's \rp~-- a large-scale online experiment in which millions of users, self-organized in thousands of communities, clashed and collaborated in an effort to realize their agenda.
    
    Our hybrid model achieves a high F1 prediction score of 0.826. We find that coarse meta-features are as important for prediction accuracy as fine-grained textual cues, while explicit structural features play a smaller role. Interpreting our model, we provide and support various social insights about the unique characteristics of the communities that participated in the \rp~experiment.
    
    Our results and analysis shed light on the complex social dynamics that drive collective behavior, and on the factors that propel user coordination. The scale and the unique conditions of the \rp~experiment suggest that our findings may apply in broader contexts, such as online activism, (countering) the spread of hate speech and reducing political polarization. The broader applicability of the model is demonstrated through an extensive analysis of the WallStreetBets community, their role in \rp~and four years later, in the GameStop short squeeze campaign of 2021. 
\end{abstract}



\maketitle

\section{Introduction}
\label{sec:intro}
Group dynamics, organization, norms, structure, and collective action are at the core of social sciences research, e.g., \cite{Lewin1947groupd,granovetter1973strength,zachary1977information,ostrom2000collective,olson2009logic,fisher2019science} to mention just a few works. The surge of online activity provides an unprecedented opportunity to study these patterns organically and at a large scale \cite{melucci1996challenging,lazer2009life}. Coordinated online activity was found to fuel street protests \cite{jackson2016ferguson,jost2018social,fisher2019science}, stimulate political conversation \cite{bail2016combining}, model the support for a social change \cite{hassler2020large}, and change traditional financial behaviours \cite{lucchini2021reddit}. The susceptibility of millions of users to emotional manipulation \cite{kramer2014experimental,brady2017emotion} and the echo-chamber effect \cite{bakshy2015exposure,del2016echo}, were utilized to disseminate misinformation, discredit democratic institutions and to interfere with political processes \cite{bovet2019influence,hanouna2019sharp,grinberg2019fake}. Notorious examples are the activity of Russian trolls \cite{jamieson2018cyberwar,benkler2018network,mueller2019report} and micro-targeting practices used by firms like Cambridge Analytica \cite{ward2018social, hu2020cambridge}. Understanding the factors that impact the behavior of online communities is a crucial step toward increased resilience of online communities to manipulation efforts \cite{basol2020good}. Despite its significance, a large-scale study of the ways decentralized communities operate, internally and with respect to other communities, is scarce (see survey in Section \ref{sec:related_work}). 

In order to model collective actions, we develop a hybrid algorithm to predict the collective action of thousands of communities, integrating multiple signals, ranging from the language-use to the community structure. 
Specifically, we use the activity of Reddit communities before the \rp~``experiment'' -- a massive online game, which we view as an external shock to the platform -- predicting the community reaction to the shock. We refer to \rp~as a naturally occurring-large scale controlled experiment. The simple design of the experiment (explained below) provides a unique opportunity to study how decentralized online communities act in response to the external shock. Our analysis illuminates the factors that facilitate the undertaking of a collective effort.

\begin{figure*}
\normalsize
    \centering
    \subfloat[{$t_0+2$ hours \label{subfig:place2hrs}}]{{\includegraphics[scale=0.112]{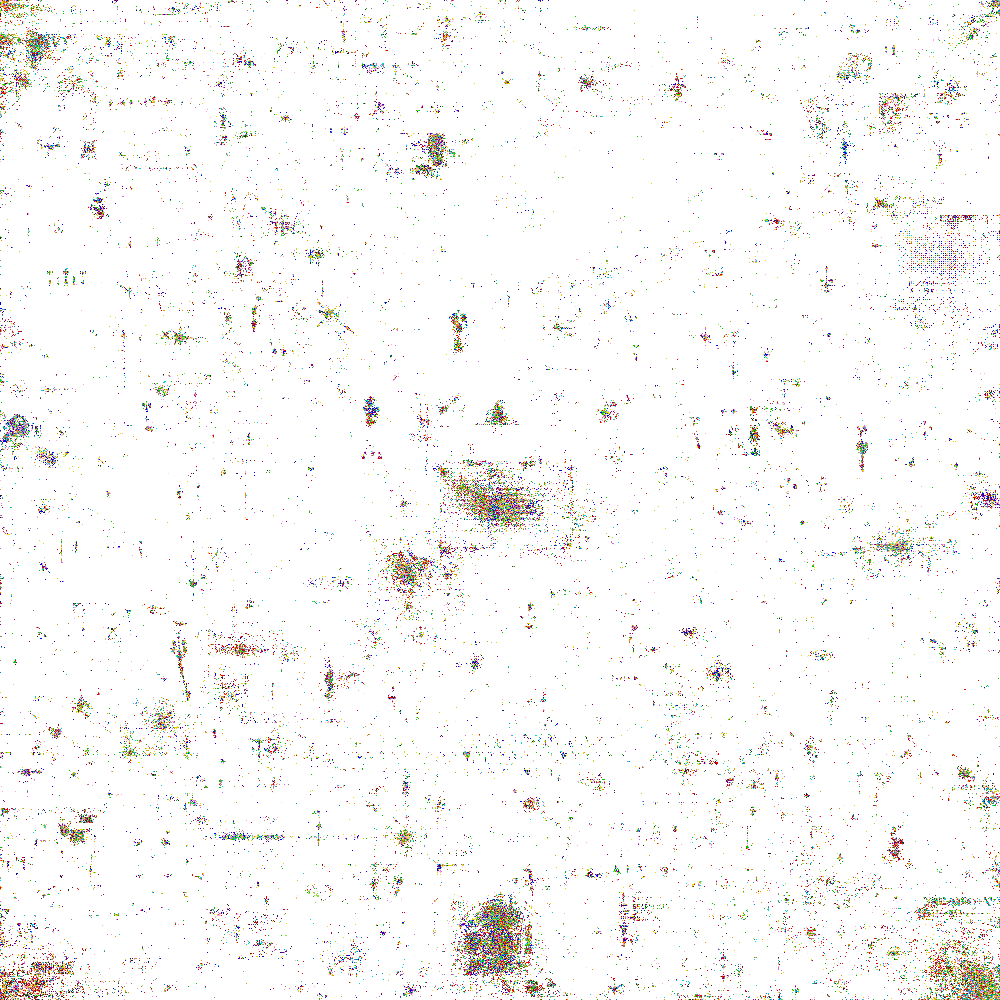}}}
    \qquad
    \subfloat[$t_0+7$ hours 
    \label{subfig:place7hrs}]{{\includegraphics[scale=0.028]{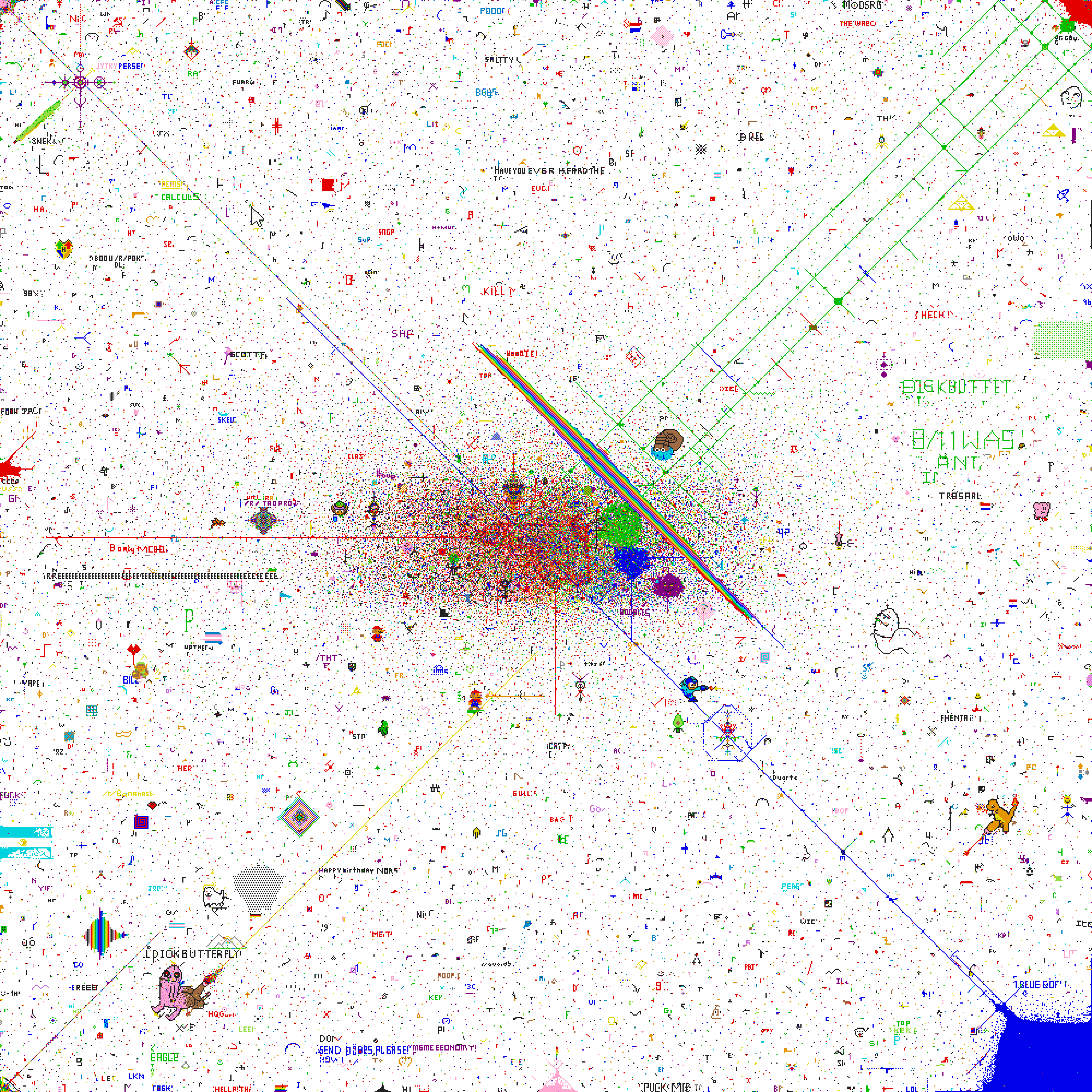}}}
    \qquad
    \subfloat[$t_0+25$ hours\label{subfig:place25hrs} ]{{\includegraphics[scale=0.028]{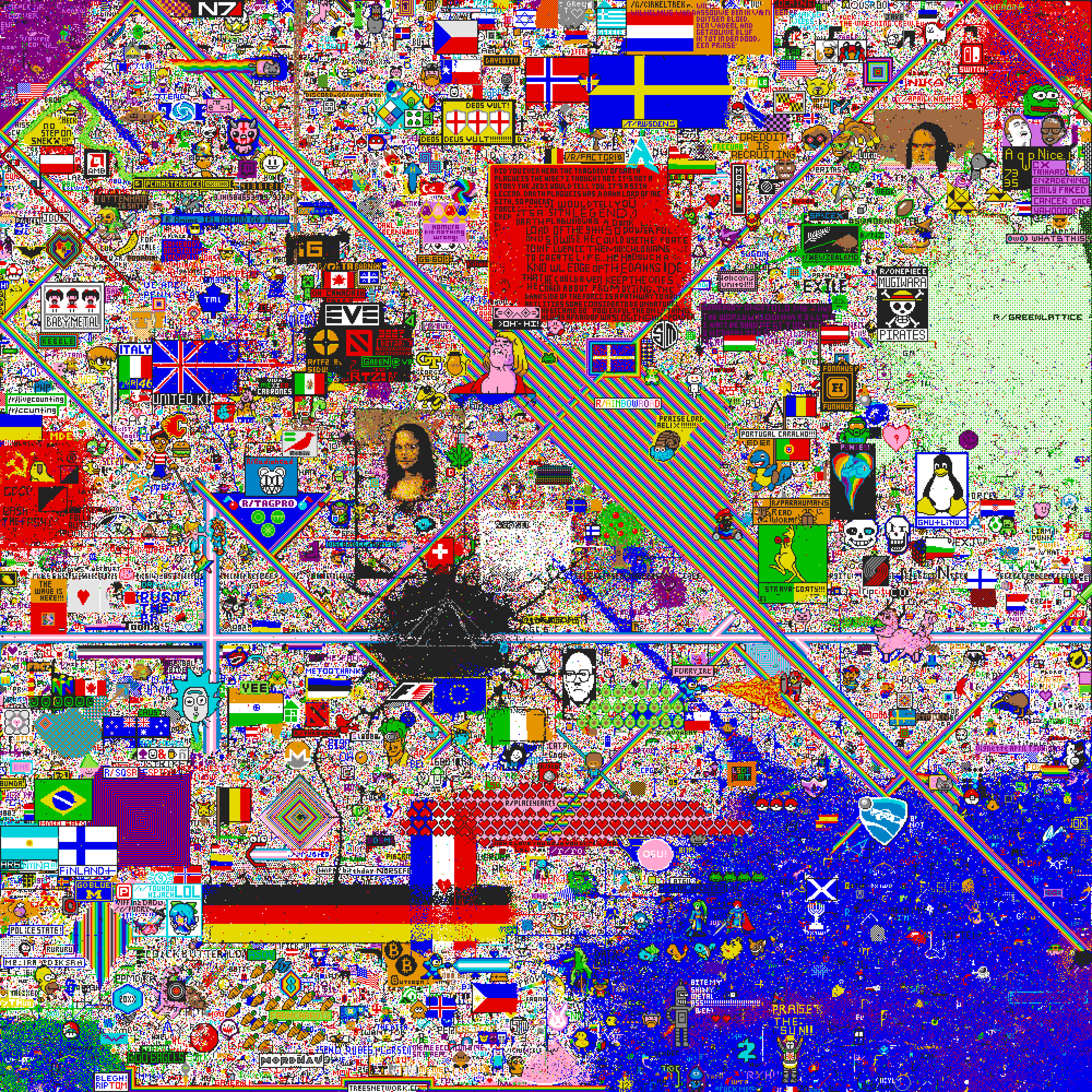}}}
    \qquad
    \subfloat[$t_0+72$ hours (final state)\label{subfig:place72hrs} ]{{\includegraphics[scale=0.028]{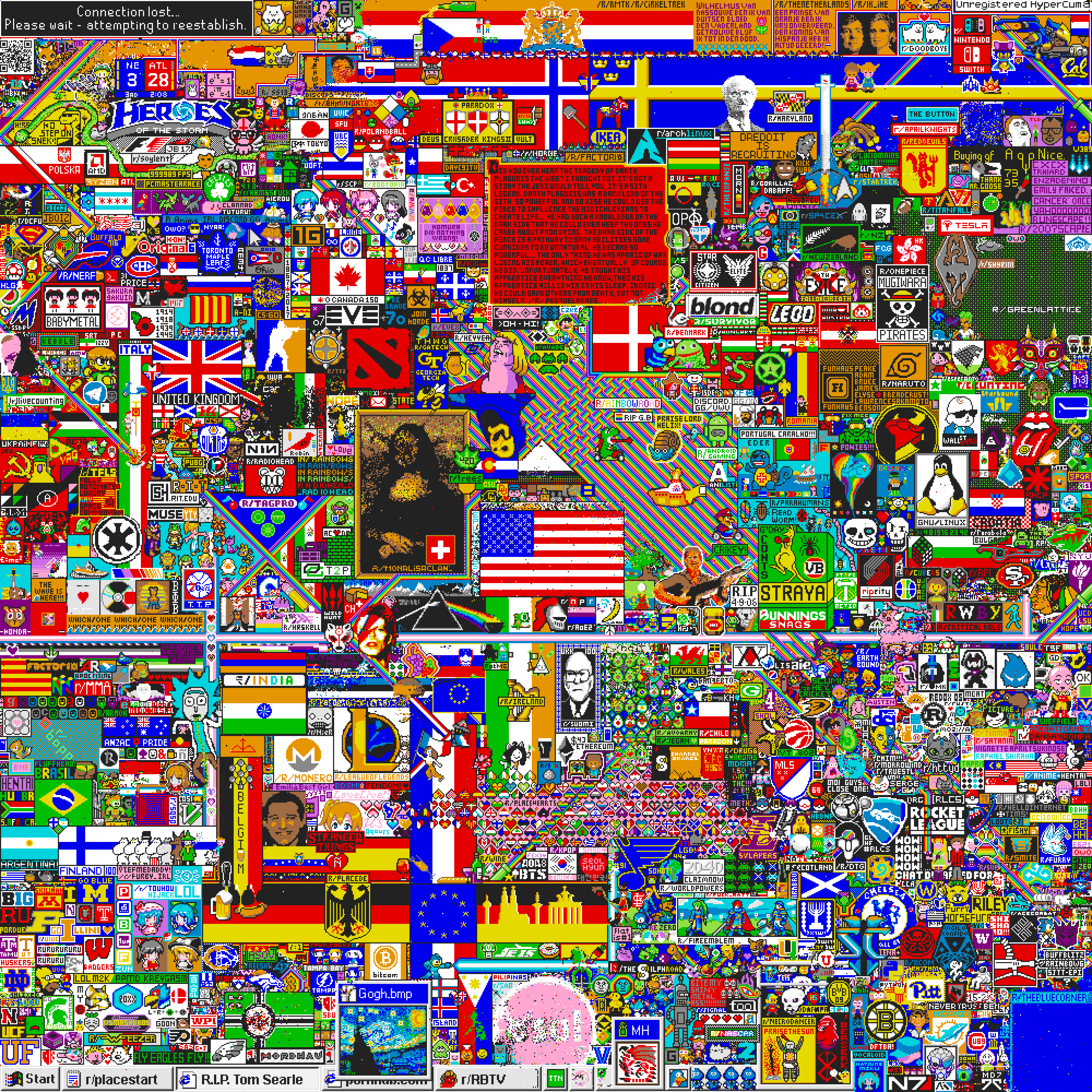}}}
    \caption{\normalsize The evolution (from left to right) of the \rp~experiment virtual canvas.}
    \label{fig:place_evolvtion}
\end{figure*}

\paragraph{{\bf Reddit}} Reddit\footnote{\href{https://www.reddit.com}{https://www.reddit.com}} is one of the most popular websites worldwide, falling short behind Google and YouTube, with an average of over 430 million active users per month \cite{reddit2019stats}. 
Reddit evolved into an active system of dedicated forums (3.1M forums, as of September 2021). Forums' URLs are marked with a 'r/' followed by the forum's name, and are therefore called \emph{subreddits}. Subreddits are usually topical, although the specificity of topics range widely, from the very broad \emph{r/music} or \emph{r/politics} subreddits to the more specific \emph{r/talkingheads} or \emph{r/brexit}. Each subreddit forms a community in which members (called \emph{redditors}) can start a new discussion thread, add comments to a thread, up/down-vote posts, etc. Each subreddit is maintained by a few moderators and has its own rules specifying community code of conduct (or lack of). In the current study, we view each subreddit as a community, having its own internal language, style, interaction dynamics, and norms.

\paragraph{{ \bf r/place -- a large-scale online ``experiment''}}
\label{subsec:rplace}
The \rp\footnote{\href{https://www.reddit.com/r/place/}{https://www.reddit.com/r/place/}} subreddit was conceived by Reddit as a social experiment gag for the 2017 April Fools’ Day. A shared white canvas of one million pixels (1000 x 1000)  appeared in a new subreddit called r/place. Redditors could change the color of any pixel, one pixel at a time. Every change of a pixel was reflected on the \emph{shared} canvas, thus viewed by all users. Viewing it as a weird online game, users were not aware of any clear purpose of the experiment. To the users' surprise, the experiment abruptly terminated after 72 hours. 
During its 72 hours of operation, the canvas attracted 16.1 million pixel changes performed by 1.2 million redditors. Figure \ref{fig:place_evolvtion} presents four snapshots attesting to the progression of the canvas' state, from its early chaotic state to its final form -- an intricate collage of complex logos and artworks. 	

The social experiment of \rp~was controlled, to some degree, as Reddit's developers served as moderators and enforced certain rules -- albeit users were oblivious to those rules.	
Two examples of such rules are: (i) Only accounts created  prior to the unexpected beginning of the experiment could manipulate the color of a pixel, and (ii) Once a redditor manipulated the color of a pixel, he or she was blocked by the system for a random time (5–20 minutes), thus effectively preventing any single redditor from having a significant influence on the canvas. 

A careful look at the final state of the canvas (Figure \ref{subfig:place72hrs}) shows\footnote{A high resolution image can be found  at \href{https://bit.ly/39e1E9a}{https://bit.ly/39e1E9a}} that many of the symbols reflect some form of group identity, including national flags, university and sports teams logos,  developer's communities (e.g., Linux), and online gaming communities. Other artworks include popular memes and reproductions of iconic art works (e.g., Da Vinci's Mona Lisa and Van Gogh's Starry Night). Some communities pursued a more abstract presence (e.g., `the blue corner', bottom right in Figures \ref{subfig:place7hrs}-\ref{subfig:place72hrs}) or the `rainbow road' which crosses the canvas diagonally. Few communities tried to drew hate-related symbols (e.g., swastikas and Pepe the Frog), or vandalized other symbols. The most successful vandalizing effort was `The-Black-Void' (TBV) -- an ever expanding black fractal-like shape in the middle of the canvas, noticeable in the center of Figure \ref{subfig:place25hrs}. It is important to note that while TBV's declared purpose was vandalizing other symbols, clashes erupted between many other communities competing for ``real estate'' for their logo. In this work we use the complex signals \emph{preceding} the experiment in order predict community engagement in the game. The rich community dynamics observed \emph{during} the experiment will be addressed in a subsequent work. 

The remainder of the paper is organized as follows: in Section \ref{sec:related_work} we provide a brief review of the relevant literature. In Section \ref{sec:data} we describe the data in detail and specify the prediction task. In Section \ref{sec:exp_setup} we present our computational approach, the various algorithms we use and the experimental setting. In Section \ref{sec:res_and_analysis} we present the results, provide a social interpretation to the results, present an error analysis, and dive into one exemplar of the \emph{r/WallStreetBets} community. Lastly, in Section \ref{sec:closure} we summarize our work and suggest future research directions.

\section{Related Work}
\label{sec:related_work}
Community behaviour has been studied for decades. \citet{Lewin1947groupd} established the modern field of group dynamics, organization, norms and collective action. Naturally, recent research puts increased emphasis on online communities \cite{lazer2009life, zhang2017community, mensah2020characterizing}.

 Many works study collaborative patterns in crowd sourced projects like Q\&A sites, Wikipedia or open-source projects \cite{keegan2011hot,keegan2012editors,ransbotham2011membership}. These works differ from ours in that such works model individuals forming a community dedicated to a project (whether a wiki page, a QA forum or an open source project), whereas we are interested in existing organic communities that are mobilized to participate in a new campaign.

A general overview of the uses of Reddit data to study its communities is presented by \citet{medvedev2017anatomy}.
Reddit communities, patterns of information sharing, and evolving community norms are studied in a series of works \cite{weninger2013exploration,de2014mental,choi2015characterizing,stoddard2015popularity,newell2016user,cunha2016effect,panek2018effects,fiesler2018reddit,DBLP:conf/icwsm/RappazC0A18, mensah2020characterizing}, among others. These works address various aspects in community organization as an interest group, the dealings with heated topics and the inherent tension between anonymity and identity. Recent works study the structure and other characteristics of Reddit communities \cite{zhang2017community, hamilton2017loyalty,kumar2018community, zhou2020condolences, massachs2020roots}. \citet{zhang2017community} suggests a new representation of communities through two complex dimensions ('distinctiveness' and 'dynamicity') and prove that these representation successfully reflects different user engagement measures (e.g., retention rate). \citet{kumar2018community} suggests a novel way to model conflict between communities on the web. Their approach combines textual features with communities meta features (all in the form of numeric vectors). This methodology is similar to the way we combine different representations of online communities. \citet{datta2019extracting} expand this work and study the landscape of conflicts among subreddits. Other works focus on a single community, presenting its uniqueness and norms \cite{jones2019r, august2020explain, britt2021oral}

Works using the rich \rp~data are still scarce. The dependency between logo size and canvas density over time is studied by \citet{muller2018compression}, while latent patterns of collaboration between individuals are modeled by \citet{DBLP:conf/icwsm/RappazC0A18} and \citet{armstrong2018coordination}. Conflicts between communities during the \rp~ campaign are studied by \cite{vachher2020understanding}. These works differ from ours in four fundamental ways: (i) These works focus on collaborations between individual redditors and the ways they correlate with other individuals, using the canvas as the main shared focal point. We, on the other hand, are interested in coordination on the \emph{community} level. Thus, we model the ways a community is organized and operates reflect on its interest and ability to undertake a competitive, large scale community effort. Consequently, (ii) We define a \emph{novel prediction task} -- which community will engage in the experiment. (iii)  We only use data \emph{preceding} the experiment while prior works use the data generated during the experiment, and (iv) We combine \emph{multiple types of features} (language, community structure, user dynamics, etc.) while others use only the \rp~pixel allocations data.
 
To the best of our knowledge, this is the first work to make use of historical Reddit data, language use and community meta features in the context of organizational development at large and in the context of the \rp~experiment in particular.

\section{Data}
\label{sec:data}
For the purpose of this research, we collected two datasets. The first, denoted $DS1$, is composed of all data posted on reddit in the six month \emph{prior} to the experiment (10/1/16--3/31/17). The second dataset, $DS2$, contains all data posted \emph{during} the 72 hours of \rp~(3/31--4/3/17) and is used to validate and improve the gold labels, as we describe below.
We also used Reddit's API\footnote{\href{https://www.reddit.com/dev/api/}{https://www.reddit.com/dev/api/}} to retrieve meta data per each subreddit (e.g., creation time, number of subscribers).

Predicting that an inactive community (i.e., by which no post was added to the platform in the months before the experiment) will not engage in the game is trivial. Hence, we created a labeled dataset, balancing the positive samples (participating communities) coupled a negative set with relatively similar attributes. The careful process of creating the balanced gold standard is described below.

\paragraph{{ \bf Participating Communities  ($S^+$)}}
The positive set is composed of the communities that took part in the experiment. While some of these communities are easy to pinpoint, others require some effort. Hence, we took two approaches toward identifying the participating communities: (i) Utilizing the place-atlas\footnote{\href{https://draemm.li/various/place-atlas}{https://draemm.li/various/place-atlas}} resource, and (ii) A machine learning approach, based on the $DS2$ dataset.

The place-atlas is a crowd-sourced effort created by a group of redditors after the termination of the experiment. This platform allowed other redditors to ``claim their victory'' by posting their community name, the location of their final artwork, and some extra details about it (e.g., some logos were the result of joint efforts by multiple communities).
Manually analyzing the atlas data we identified 802 communities participating in \rp. However, most of the artworks indexed in the place-atlas are attributed to `winners' -- communities whose effort is recognized on the canvas at the termination of the experiment. Many other communities did participate, but failed to leave a mark, as their logo was vandalized or over-ridden by a competing community. In order to identify these, as well as other communities that were not indexed in the place-atlas, we trained a simple classification model based on regular expressions. 

We used the data that were generated \emph{during} the 72 hours of the experiment ($DS2$) matching specific regular expressions that could link communities to the experiment (e.g., \emph{draw}, \emph{rplace}, \emph{pixel}, \emph{canvas}). All matched subreddits are \emph{candidates} for the positive set. This pool of candidates needs to be filtered further since many communities only discuss the sudden \rp~ hype but do not end up participating. We thus manually labeled \textasciitilde 100 communities as positive (drawing) and another \textasciitilde 100 communities as negative (not drawing). This way we had a small labeled seed to be used in a bootstrapping manner, discovering more participating communities with each iteration. Following \citet{kozareva2010not}, only candidates with a very high confidence score were added to the seed in each iteration. After three iterations we discovered a few hundreds of new positive examples (i.e., communities that were not indexed in the place-atlas). We manually validated\footnote{By reading the posts that were written in the subreddit during the time of the experiment. We looked for explicit statements by the subreddit members describing a joint effort of participation in \rp.} a sample of these newly found positive examples, and found it highly accurate ($94.9\%$ accuracy over 156 communities). Adding the discovered positive communities to those that were identified by the place-atlas, resulted in a set of 1231 communities. We refer to this set of communities as $S^+$.

\begin{table*}
\centering
\caption{\small Subreddits statistics for $S^+$ (drawing), $S^-$ (not drawing) and $S$  (all the subreddits active in the 6-month period prior to the \rp~ experiment). Mean values are computed over all submissions in a subreddit. \emph{Age} denotes the number of days since the subreddit was established. Inactivity denotes the number of days from the most recent post to the launch of  \rp~(effectively bound by 180 -- the span of $DS1$).}
{\normalsize 
    \begin{tabular}{l@{ }|c@{\quad}c@{\quad}c@{\quad}c@{\quad}|c@{\quad}c@{\quad}c@{\quad}c@{\quad}|c@{\quad}c@{\quad}c@{\quad}c@{}}
      \multicolumn{1}{c}{} & \multicolumn{4}{c}{\boldmath{$S^+$}\textbf{(1231)}} & \multicolumn{4}{c}{\boldmath{$S^-$}\textbf{(1289)}} &
      \multicolumn{4}{c}{\boldmath{$S$} \textbf{(258.8K)}} \\[2pt]
      {} & {Total} & {Mean} & {Median} & {STD} & {Total} & {Mean} & {Median} & {STD} & {Total} & {Mean} & {Median} & {STD} \\[2pt]
      \hline\rule{0pt}{12pt}
      Subscribers	    &219.7M  &178.5K   &24.1K  &1.13M  &409.2M   &317.5K    &15.9K    &1.9M  &1.7B   &6.6K    &30.0K    &234.6K\\[2pt]
    Active Users	    &6.9M  &5.6K   &1.47K  &15.9K  &4.4M   &3.47K    &0.7K    &12K  &54.7M  &211.4   &2.0  &5.9K\\[2pt]
      Age	&--  &2.15K   &2.34K  &0.89K  &--   &1.7K    &1.7K    &1.01K &--  &0.98K   &0.87K  &0.8K\\[2pt]
      Inactive    &--  &0.9   &0.0  &6.7  &--   &2.23    &0.0    &11.1 &--  &55.2   &39.0  &52.74\\[2pt]
      Submissions	    &12.8M  &10.4K   &2.3K  &55.6K  &12.3M   &9.6K    &2.4K    &51.7K &53.3M  &209.1   &3.0  &6.1K\\[2pt]
     {\vtop{\hbox{\strut Comments}}}	&--  &12.0   &8.6  &25.3  &--   &6.1    &3.9    &7.5  &--  &1.4   &0.25  &23.9\\[2pt]
    \hline
    \end{tabular}
   \label{table:Data_statistics}
}
\end{table*}

\paragraph{{\bf Non-participating communities ($S^-$)}} Most sureddits did not take an organized part in the \rp~experiment. Using all subreddits not in $S^+$ as the negative set is inappropriate for several reasons, ranging from the relative simplicity of classifying an extremely unbalanced datasets (\textasciitilde1.2K positive vs. \textasciitilde1.2M), to the fact that many subreddits are old, inactive, very new or very small (in terms of active members). In order to create a balanced and challenging dataset we opted to subsampling -- trying to have the meta features of communities in the negative set similar to those in the positive set.
The negative training set, denoted $S^-$, was therefore created in the following heuristic manner: we first matched each drawing community in $S^+$ to a non-drawing community of similar size where size is measured by: (i) Number of subscribers, or (ii) Number of submissions posted in the community page. Both measures have pros and cons, so we experimented with both. Both heuristics ended up showing very similar results. In this paper we only report results obtained from the second heuristic.\\

General statistics describing various aspects of the datasets are presented in Table \ref{table:Data_statistics}. Calculations are based on the $DS1$ (historical) dataset. Many subreddits(\textasciitilde950K) did not show any activity in the six month span of $DS1$, thus are not accounted for in the table. For example, the table shows that communities in $S^+$ and $S^-$  are bigger and more active compared to a random Reddit community. 

We wish to reiterate that $DS2$ is only used for the creation of the gold standard (participating/non-participating) feature. In the remainder of the paper we use \emph{only} the ``historical'' data ($DS1$) in representing and predicting community participation.

\subsection{Community Representation}
\label{sec:characterizing_subreddits}
Communities are multifaceted and could be characterized from different perspectives. To this end, we represent Reddit communities by features of three general types: (i) Textual features, (ii) Meta-features, and (iii) Network features.

\paragraph{{\bf Textual features}} We normalized the data by lower casing, tokenization, removing punctuation and conversion of full URL addresses to the domain name (e.g., www.youtube.com/XYZ $\rightarrow$ www.youtube.com). \ignore{Since submissions include a submission header and a submission text body (unlike comments/replies that contain only the text body) we concatenated these two parts into one.}
We used both classic Bag-of-Words (BOW) representation for the GBT model (see Section \ref{sec:exp_setup}) and word embeddings representation for the neural models.
In the BOW models, we used the {\em TF-IDF} score \cite{salton1986introduction}. Using bigrams/trigrams did not yield any improvement so all BOW results are reported for the unigram setting, using the 300 most frequent tokens.

\paragraph{{\bf  Meta-features}} Reddit was originally conceived as an interest-based message board, therefore a subreddit can be represented by meta features like the number of users subscribed to a community, the average number of posts per day, the average number of votes per post, the ``age'' of the community (days since its creation), etc. We used a total of 25 meta-features for each subreddit.

\begin{figure}%
    \centering
    \subfloat[{\emph{r/srilanka} subreddit}]{{\includegraphics[width=3.3cm]{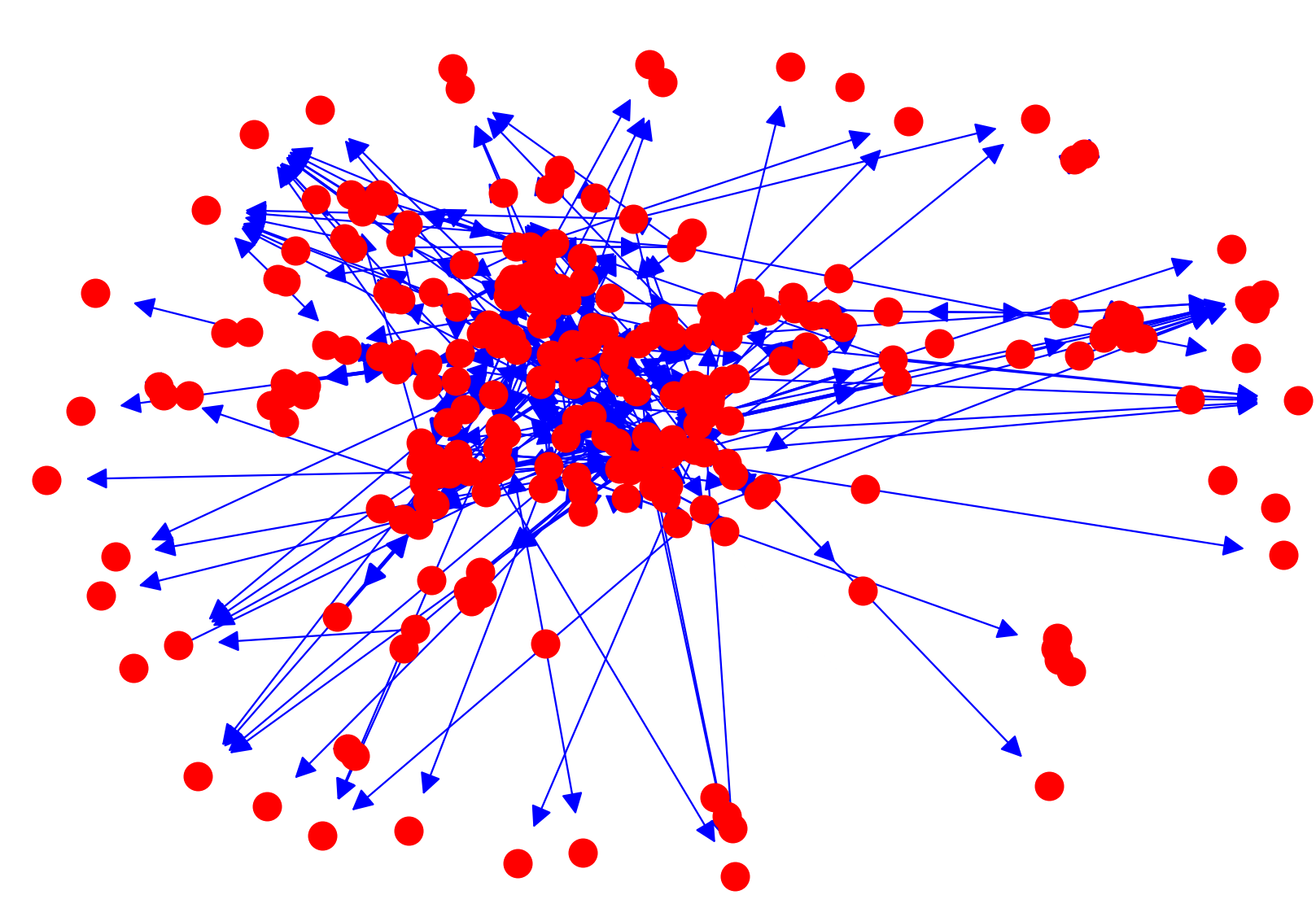} }}
    \qquad
    \subfloat[\emph{r/insults} subreddit]{{\includegraphics[width=3.3cm]{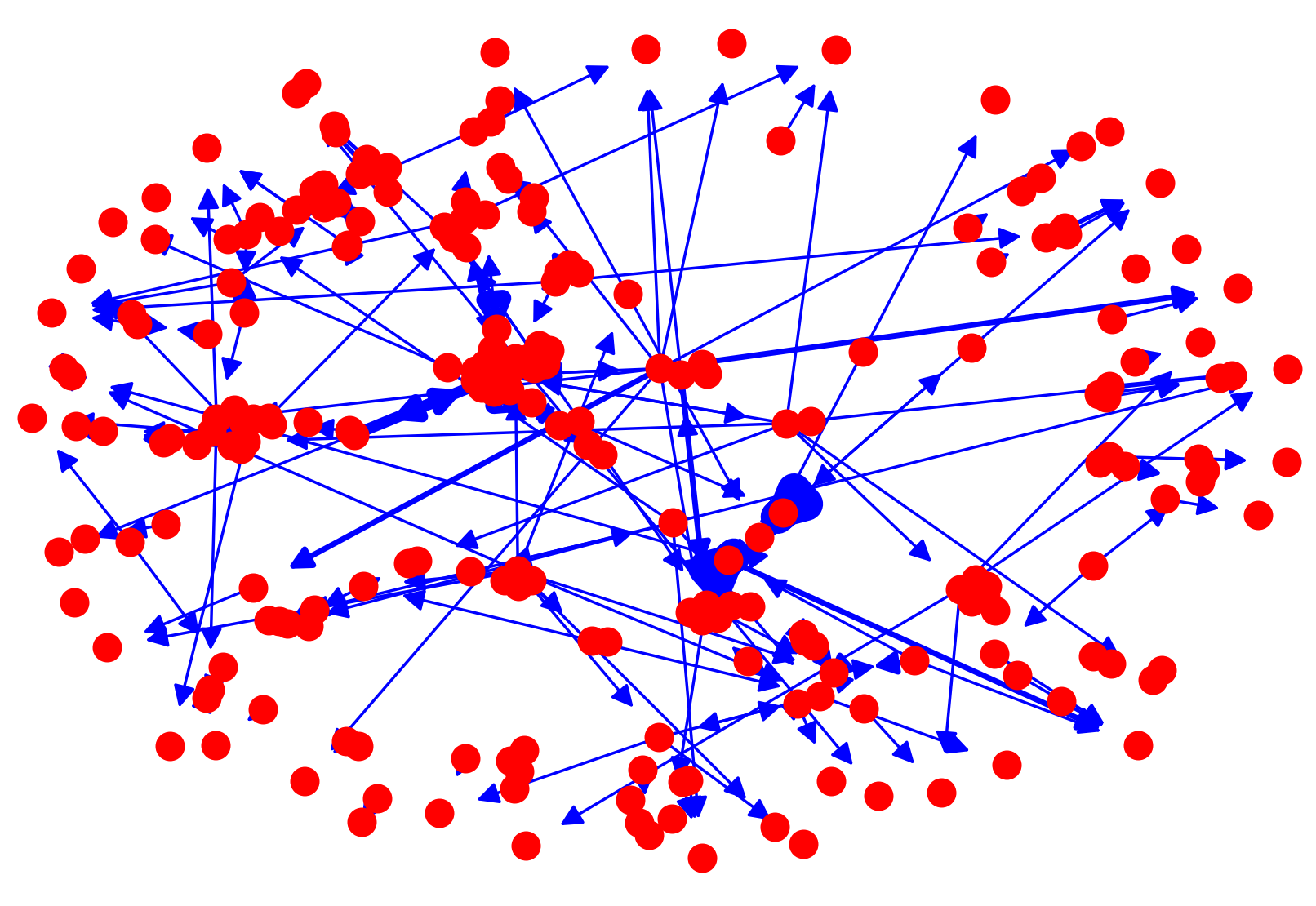} }}
    \caption{\normalsize Network structure of two communities with a similar number of nodes. Both networks were rendered using the Fruchterman-Reingold layout.}
    \label{fig:network_example}
\end{figure}

\paragraph{{\bf  Network features}} A community can be characterized by the patterns of communication between its members. These interaction patterns could be thought of as a social network in which a direct reply by user $u$ to a post by user $v$ constitutes a directed edge $u \rightarrow v$. These networks provide another perspective on the organizational principles of a community and the dynamics between its members.
The intuition behind the use of the network perspective is illustrated through Figure \ref{fig:network_example}. The figure shows the network structure of two subreddit communities with a similar number of nodes: \emph{r/srilanka} ($\in S^+$) and \emph{r/insults} ($\in S^-$). It is visually apparent that these communities are organized in a very different way with \emph{r/srilanka} presenting a tighter community structure. 
In total, we used 32 network statistics as features (e.g., \#nodes, \#edges, avg. and std. of various centrality measures, \#triangles)\footnote{The list of all meta and network features, together with a short description of each, is provided in Table \ref{table:meta_and_network_feautures_description} in Appendix \ref{sec:appendix_meta_and_network_features}.}

\section{Experimental Setting}
\label{sec:exp_setup}
We cast the likelihood of a community to participate in \rp~as a classification problem. We experiment with an array of algorithms ranging from a logistic regression based on bag of words to a neural architectures, taking a collection of complex features as input. The algorithms we experiment with could be divided into two broad categories: sequential and non-sequential models. Sequential models take the sequential nature of the data (text, discussion threads) into account while non-sequential models ignore this structure. We consider a number of classifiers of each type. However, due to space constraints, we report only on the following five: Gradient Boosted Decision Trees (GBT), Multi-Layer Perceptron (MLP), Zero-Shot BERT with an MLP layer, Parallel LSTM with an MLP layer, and a Max-Pooling CNN with an MLP layer. A deviance loss function is used for the GBT algorithm and a binary log-loss is used for the MLP and the parallel LSTM models. We provide a broader explanation for each classifier in Appendix \ref{sec:appendix_alg_approaches}.
Each subreddit is represented by an array of features extracted from the ``history'' of the community ($DS1$) -- meta features, textual features, and network features (see Section \ref{sec:data}).

We execute all algorithms in an ablation manner, in order to learn the importance of the different feature types. Accuracy, Precision, Recall, and F1-score are reported for each setting.  The different algorithms are optimizing the F1-score as precision and recall are equally important, given the task definition.
We evaluate all algorithms and settings using a stratified 5-fold cross validation. Neural architectures are restricted to a maximum of ten epochs with an early stopping.

For the MLP implementation, we apply a simple neural model -- a single hidden layer of 150 nodes. For the parallel LSTM deep neural model (see Figure \ref{fig:parallel_lstm} in Appendix \ref{sec:appendix_alg_approaches}), we setup 150 hidden units in each LSTM hidden state.

We use the sklearn \cite{pedregosa2011scikit} implementation of the GBT algorithm. We use the DyNet \cite{neubig2017dynet} and PyTorch packages \cite{paszke2019pytorch} for building the neural architectures, due to the dynamic construction of the computation graph which is highly efficient given the high variance in number and length of textual data. We make the code developed as part of the research and all data that were collected public in our project's GitHub repository.\footnote{\href{https://github.com/NasLabBgu/rplace-engagement-prediction}{https://github.com/NasLabBgu/rplace-engagement-prediction}}

\begin{table}
\centering
\caption{\small Prediction results. L/M/N denote \emph{linguistic}, \emph{meta} and \emph{network} features, respectively. The Zero-shot BERT, parallel LSTM, and parallel CNN models require linguistic features in all settings, hence only three options were evaluated for these models. 
}
    {\small    
    \begin{tabular}{l@{\quad}|l@{\quad}|l@{\quad}@{\quad}c@{\quad}c@{\quad}c@{\quad}c@{\quad}}
      \multicolumn{6}{c}{}\\[2pt]
      {Model} & {Features} & {Precision} & {Recall} & {F1-Score}\\[2pt]
      \hline\rule{0pt}{12pt}
    \multirow{5}{*}{\vtop{\hbox{\strut GBT}\hbox{\strut Models}}}
      & L &0.776$\pm$0.027		&0.758$\pm$0.033  &0.766$\pm$0.022\\[2pt]

      & M &0.766$\pm$0.027		&0.8$\pm$0.039  &0.782$\pm$0.017\\[2pt]

       & N &0.67 $\pm$0.044		&0.75$\pm$0.031  &0.707$\pm$0.027\\[2pt]

       & M + N  &0.765$\pm$0.062		&0.814$\pm$0.031  &0.788$\pm$0.011\\[2pt]

        & L + M &0.8$\pm$0.024		&0.82$\pm$0.027 &0.81$\pm$0.016\\[2pt]

       & L + M + N  &0.814$\pm$0.03		&\textbf{0.84$\pm$0.018} &\textbf{0.826$\pm$0.011}\\[2pt]
    \hline\rule{0pt}{12pt}
    \multirow{6}{*}{\vtop{\hbox{\strut MLP}\hbox{\strut Models}}}
        & L &0.76$\pm$0.049		&0.74$\pm$0.063    &0.74$\pm$0.017\\[2pt]

        & M &0.737$\pm$0.033		        &0.632$\pm$0.031    &0.679$\pm$0.013\\[2pt]

        & N &0.68$\pm$0.054	   &0.501$\pm$0.134    &0.57$\pm$0.083\\[2pt]

        & M + N &0.73$\pm$0.028		&0.76$\pm$0.067     &0.77$\pm$0.02\\[2pt]

        & L + M &0.78$\pm$0.039		&0.82$\pm$0.064     &0.784$\pm$0.021\\[2pt]

        & L + M + N &0.755$\pm$0.056		&0.82$\pm$0.006     &0.78$\pm$0.021\\[2pt]
        \hline\rule{0pt}{12pt}
     \multirow{3}{*}{\vtop{\hbox{\strut Zero-Shot}\hbox{\strut BERT}\hbox{\strut Models}}}
        & L
        &0.582$\pm$0.006     &0.825$\pm$0.085 &   0.681$\pm$0.028\\[2pt]

        & L + M
        & 0.697$\pm$0.036    &0.806$\pm$0.108     &0.744$\pm$0.047\\[2pt]

        & L + M + N
        &0.761$\pm$0.062     &0.686$\pm$0.128 &0.713$\pm$0.051\\[2pt]
    \hline\rule{0pt}{12pt}
    \multirow{3}{*}{\vtop{\hbox{\strut Parallel}\hbox{\strut LSTM}\hbox{\strut Models}}}
        & L &0.741$\pm$0.06		&0.752$\pm$0.085 &0.741$\pm$0.03\\[2pt]

        & L + M
        &0.743$\pm$0.024  &0.799$\pm$0.07 &0.767$\pm$0.011\\[2pt]

        & L + M + N  &0.779$\pm$0.04	&0.756$\pm$0.089 &0.763$\pm$0.032\\[2pt]
    \hline\rule{0pt}{12pt}
    \multirow{3}{*}{\vtop{\hbox{\strut Parallel}\hbox{\strut CNN}\hbox{\strut Models}}}
        & L
        &0.761$\pm$0.019     &0.722$\pm$0.08 &   0.737$\pm$0.037\\[2pt]

        & L + M
        & 0.795$\pm$0.012    &0.726$\pm$0.062     &0.756$\pm$0.021\\[2pt]

        & L + M + N
        &\textbf{0.825$\pm$0.02}     &0.753$\pm$0.042 &0.786$\pm$0.021\\[2pt]
    \end{tabular}
   \label{table:results}
}
\end{table}

\section{Results and Analysis}
\label{sec:res_and_analysis}
\subsection{Prediction Accuracy}
The results obtained by each model in the various ablation settings are detailed in Table \ref{table:results}. The parallel CNN neural model, using all feature types, achieved the best precision score ($0.825$). However, the best F1-score ($0.826$) was obtained by combining all feature types in the GBT model. Using each feature type independently performs well, yielding a significant improvement compared to a random baseline. The explicit network features play a lesser role compared to other feature types.

Using a single feature type, the highest F-score is obtained by the community meta features (GBT), beating the language features by a small but a significant margin. However, this holds only for the GBT model, probably due to the power of the word embeddings used in the neural models. We note that while trailing behind, network features alone achieved a decent ($\thicksim20\%$) improvement over a random baseline. These results show that all feature types capture a signal inherent to the community's participation in the experiment, whether linguistically or structurally. 

Looking at ablation tests of the best performing algorithm, combining all three feature types achieves superior results over all subsets of features. Surprisingly though, three of the neural models (i.e., the MLP, the Zero-Shot BERT, and the Parallel LSTM) performed better without the network features. This does not hold for the CNN -- the best performing neural architecture we tried. These results suggest that while the network features are instrumental in improving the classification, this feature type could benefit from further tuning.

As noted, the GBT model performs better than the neural models across all settings. This result demonstrates the often overlooked limitations of neural networks and language models \cite{merrill2021provable}. Specifically, we attribute this result to the discrepancy between the relatively small number of instances in each class and the richness of the feature types. We elaborate on that in Section \ref{subsec:performance_of_neural_models}.  

\subsection{Error Analysis}
\label{subsec:error_analysis}
We further conducted an error analysis, focusing on the false positive/negative predictions of the best performing model.

Table \ref{table:model_errors} contains the most striking errors of the GBT model. Interestingly, many of the false negative cases are `ImagesOf\textless geo-location\textgreater' subreddits. While these communities have a strong and defined identity, related to the specific location in focus, they present patterns of general picture-posting communities rather than those of an active community that promote engagement between its members. For example, while participating in \rp, joining efforts with the \emph{r/texas} community, the \emph{r/ImagesOfTexas} subreddit presents dynamics closer to those of \emph{r/catPics}. In that sense, \emph{r/ImagesOfTexas} can be seen as a sub-subreddit of \emph{r/texas}, a fact that our models fail to account for.

On the other hand, the false positive list contains some communities that are, indeed, expected (albeit naively) to participate in \rp. Gaming communities that were correctly predicted to participate (e.g., \emph{r/PuzzleAndDragons}, \emph{r/ScottPilgrim}, and \emph{r/ClashRoyale}) are very active, presenting high degree of inter-user engagement. It is therefore understandable why other communities that revolve around video games (e.g., \emph{r/mega64},  \emph{r/Dirtybomb}), bearing strong similarity to gaming communities, 
would be falsely predicted to participate in \rp.

\subsection{Social Interpretation}
\label{subsec:feature_importance}
We use the SHAP framework \cite{lundberg2017unified} to quantify the impact of specific features on the prediction and derive social interpretation and insights. A SHAP value is calculated for each explanatory variable (feature) and input instance (i.e., a community in our case) for a specific model. High (low) SHAP value  indicates the positive (negative) impact of the feature on the prediction of the specific instance. Looking at the aggregate values for a specific feature provides a way to interpret a given model.

\begin{table}[]
\centering
\small
\caption{\small The ten false positive/negative errors with the highest deviation from the expected class score. Communities in each type are ordered according to the likelihood that the GBT model assigns the community to participate in \rp~(values in brackets).}
{
    \begin{tabular}{l@{\quad}|l@{\quad}}
     \multicolumn{1}{c|}{False Positives} &
     \multicolumn{1}{c}{False Negatives} \\[2pt]
      \hline\rule{0pt}{6pt}
        r/MemeEconomy (0.97)& r/ImagesOfTennessee (0.03)\\[2pt]
        r/mega64 (0.94) & r/ImagesOfNewZealand (0.04)\\[2pt]
        r/KillLaKill (0.93) & r/RealGirls (0.05)\\[2pt]
        r/SubredditDrama (0.93) & r/ImagesOfColorado (0.05)\\[2pt]
        r/nomanshigh (0.93) & r/Pigifs (0.06)\\[2pt]
        r/MST3K (0.93) & r/thetreesnetwork (0.06)\\[2pt]
        r/Dirtybomb (0.92) & r/ImagesOfTexas (0.07)\\[2pt]
        r/totalwar (0.92) & r/srpska (0.07)\\[2pt]
        r/Battlefield (0.92) & r/BDSMcommunity (0.09)\\[2pt]
        r/ericprydz (0.92) & r/the\_donald\_discuss (0.1)\\[2pt]
    \hline
    \end{tabular}
    \label{table:model_errors}
}
\end{table}

\begin{figure*}
  \centering
    \begin{tabular}{c@{}}
        (a) Meta Features\\[2pt]
        \includegraphics[scale=0.72]{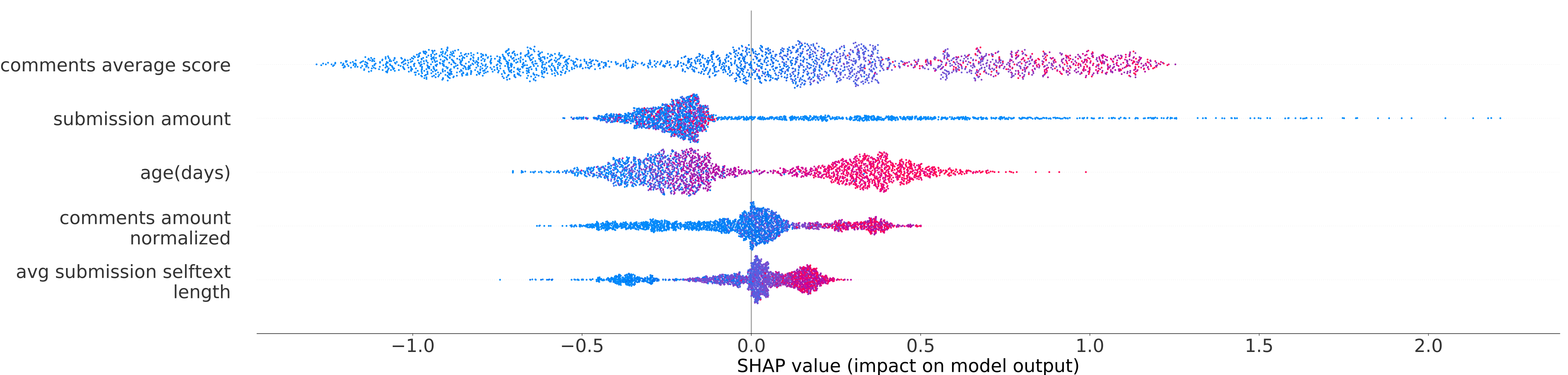}
    \end{tabular}\\
    \vspace{1cm}
    \begin{tabular}{c@{}}
        (b) Network Features\\[2pt]
        \includegraphics[scale=0.72]{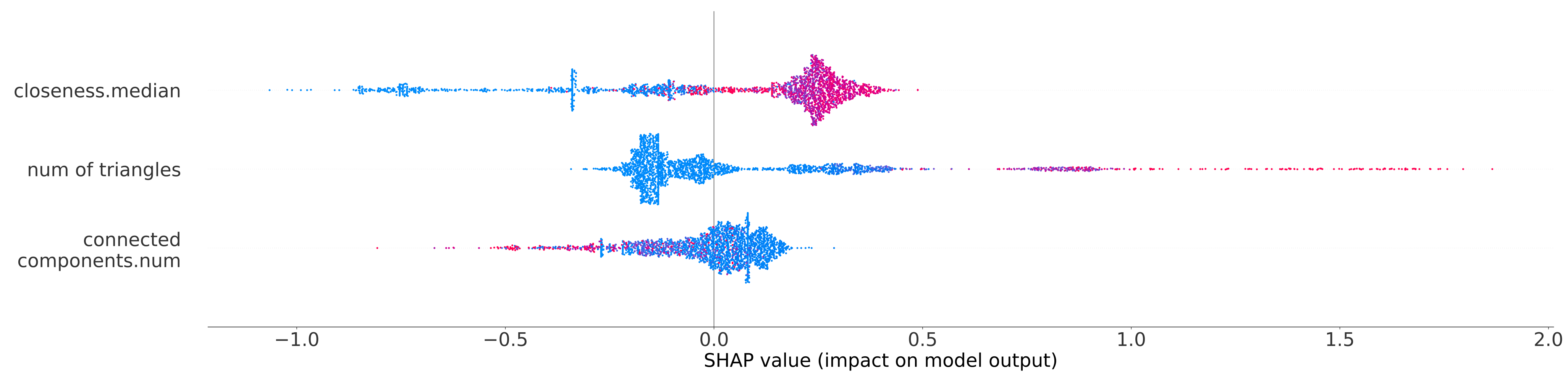} \\[\abovecaptionskip]
    \end{tabular}\\
    \vspace{0.5cm}
    \begin{tabular}{c@{}}
        (c) Textual Features\\[2pt]
        \includegraphics[scale=0.92]{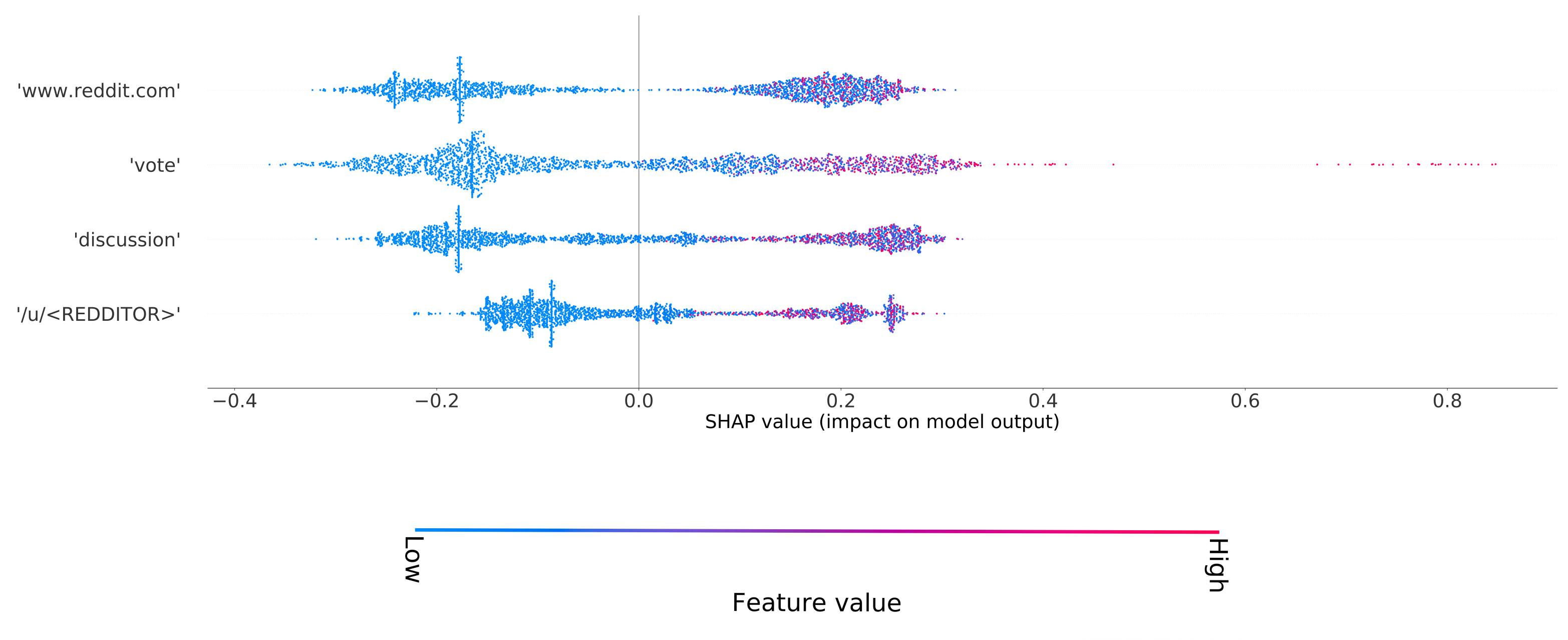} \\[\abovecaptionskip]
    \end{tabular}
\vspace{-0.6cm}
  \caption{\small SHAP values of twelve prominent features. Each point represents the SHAP value of an instance for a specific feature (in the GBT model). All features in the figure are significantly important (p-value $< 10^{-4}$)  in the prediction model.}
  \label{fig:shap_values}
\end{figure*}

The SHAP values of the most prominent features (of each feature type) are presented in Figure \ref{fig:shap_values}.
The SHAP value (X-axis) indicates the contribution of the feature to the prediction --  high value means positive contribution to the model's prediction. The Y-axis simply indicates the density of a specific SHAP value. The color corresponds to the actual value of the feature for each instance -- high values in red and low values in blue.

\subsubsection{Community Meta Features}
\paragraph{{\bf Feedback}} Feedback, particularly a positive feedback, is known to have a positive influence on the cohesiveness of communities and to increases engagement and retention \cite{cheng2014community,glenski2017predicting}. Reddit provides its users with a feedback mechanism -- users can up/down-vote submissions made by other users. We find the `comment average score' to be the most powerful feature among the meta features (see Figure \ref{fig:shap_values}a). The \emph{comment average score} per community for $s \in S^+$ is 4.39, and only 3.05 for  $s \in S^-$  -- suggesting that positive feedback (more up-votes than down-votes) supports a positive atmosphere, increases solidarity, and contributes to the community preparedness to engage in a campaign. 

\paragraph{{\bf Age}} The SHAP values of the {\em age} meta-feature are clearly distributed between two distinct clusters (Figure \ref{fig:shap_values}a). These clusters appear to be highly correlated with the actual value of this feature, as indicated by color: high SHAP values are red, and low SHAP values are blue. This supports our intuition that the longer history an active community shares, the more it will be inclined to join a distributed campaign, while a shorter shared history (blue) has a negative (low SHAP values) impact on the community's inclination to join the cause.

\subsubsection{Network Features}
\paragraph{{\bf Closeness Centrality}}  The median nodes' \emph{closeness} is significantly higher in participating communities (note that higher values of closeness indicate higher centrality, Figure \ref{fig:shap_values}b). This is an indication that participating communities are denser, maintaining tighter relations between members. 

\paragraph{{\bf Triads}} 
Communities participating in \rp~have a significantly higher number of triads. On average, participating communities have 148.46K triads compared with only 26.8K triads in non-participating ones. While this could be a attributed to the fact that $S^+$ communities tend to have a larger number active users (Table \ref{table:Data_statistics}), it is important to note that of 23\%  of $S^-$ communities had \emph{no} triads at all, compared with only 4.95\% in $S^+$.
Indeed, community cohesion is known to impact the inclination of its members to undertake a collective task \cite{friggeri2011triangles, fagnan2014using}.

\paragraph{{\bf Connected Components}}
The number of connected components is negatively correlated with the SHAP values, suggesting that fractured communities are less likely to engage in a collective action. On average, participating communities have 46.79 connected components, while non-participating ones have 69.32.
While this result seems trivial (group cohesiveness is a major factor in acting toward a common goal), the relative cohesiveness of the participating communities deviates from the expected number of components given community size. Considering only the size of the community (nodes amount), the number of connected components is expected to positively correlate with the number of nodes in the network. However, in the \rp~case, we observe the opposite -- participating communities ($S^+$) have a significantly \emph{higher} number of active users\footnote{Note that the number of subscribers (first row in Table \ref{table:Data_statistics}) is less relevant when analysing connected components as we only consider users that took an active role in their community in the months preceding the experiment.}, compared to $S^-$ communities (average and median, see Table \ref{table:Data_statistics} second row) but still present a much \emph{higher} connectivity.

\subsubsection{Textual Features}
\paragraph{{\bf Practicing Good Citizenship}} Higher {\em tf-idf} values (purple and red) for words like {\em vote} and {\em discussion} correlate with higher SHAP values (see Figure \ref{fig:shap_values}c). This suggests that communities that opted to participate in \rp~are those who tend to promote discussions among their members (rather than serve merely as message boards) and encourage their members to engage  and practice ``good citizenship'', evident by a high frequency of tokens like  \emph{vote} and \emph{discussion}. 

\paragraph{{\bf Explicit Mentions}} Members of participating communities tend to explicitly mention other users (using the `/u/' prefix), significantly more than members of non-participating communities. This is in line with well established social theory i.e., the strong correlation between the \emph{sense of community} to social factors such as knowing your neighbor names \cite{chavis2002sense}.

\paragraph{{\bf Internal URLs}} Interestingly, referencing Reddit, indicated by the {\em www.reddit.com}, also correlates with high SHAP values. We hypothesize that frequent references to Reddit indicate strong engagement with the platform and the community. 


\subsection{Beyond \rp: WallStreetBets (WSB)}
\label{subsec:wsb}
The GameStop short squeeze of early 2021, organized and promoted in the \emph{WallStreetBets} (WSB) subreddit,\footnote{A short squeeze campaign that incurred loses of billions to a number of hedge-funds in just a few days during January and February of 2021.} was argued, at the time, to have shifted the financial power balance. The unfolding of the events initiated a debate among economists and sociologists, trying to understand its causes and its impact on future trade \cite{di2021gamestop,long2021, lucchini2021reddit}. This successful short squeeze campaign involved a tight, though decentralized, coordination toward the realization of a common goal. 
Similarly to the \rp~sudden appearance, the GameStop short squeeze provides another unique, though anecdotal, opportunity to examine the way community characteristics (norms, textual, structural, etc.) correspond to a collective undertaking.

The \emph{WSB} community took part in \rp. Its participation in the experiment is somewhat surprising. Most of the communities that participated, were naturally gathered around a national flag, a video game logo, a sport team symbol, etc. These are natural causes for participation -- the community members share solidarity at a basic level and have a predefined logo (insignia) to place on the canvas. This solidarity is not expected in a community like \emph{WSB} that is used for sharing stock-trading advice. 
In spite of our early disposition regarding the nature of \emph{WSB} and similar communities, we observe the community ability to collaborate as reflected in \rp, leaving a clear mark of a sit size on the canvas\footnote{See the right side of the final canvas, Figure \ref{subfig:place72hrs} -- above the GNU/Linux penguin.}.
Indeed, keeping \emph{WSB} out of the training set, the model predicts the participation of the community with a likelihood of 0.82. \ignore{This demonstrates the power of the model to predict community behaviour based on various signals characterizing the community. }

\begin{figure}
  \centering
    \begin{tabular}{c@{}}
        \includegraphics[scale=0.33]{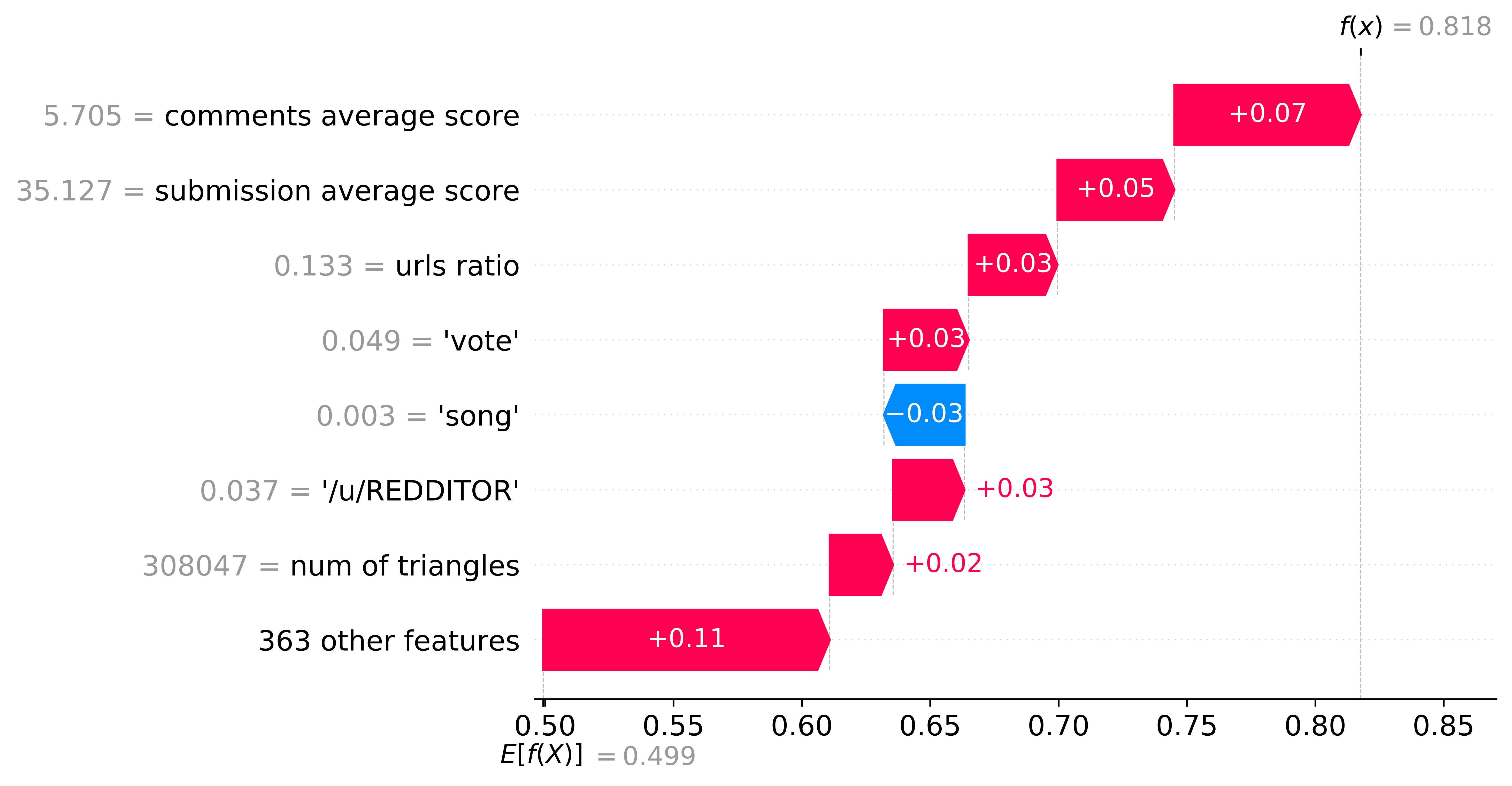}
    \end{tabular}
  \caption{\small SHAP analysis - \emph{WallStreetBets} subreddit. The top features that contribute to push the model output from the base likelihood value (0.499) to the its final value (0.818). Features pushing the prediction higher (lower) are shown in the right-red (left-blue) arrows. Values on the arrows are the marginal contribution of each feature. Values on the y-axis are the actual values of each feature.}
  \label{fig:shap_wsb_sr}
\end{figure}

We use a SHAP ``waterfall'' plot to further analysis of the community. The top features contributing to the model's prediction are presented in Figure \ref{fig:shap_wsb_sr}.
The `comments average score` is the most dominant feature for this community. All three feature types (linguistic, meta, and network) appear among the most dominant features. We observe that most of the features in Figure \ref{fig:shap_wsb_sr} are also found to be central in the \emph{aggregated} SHAP analysis (Figure \ref{fig:shap_values}). However, both `submissions average score' and `urls ratio' are ranked higher in \emph{WSB} compared with the aggregated SHAP analysis \ignore{(i.e., these features are ranked in the 6th and 7th place respectively in the aggregated SHAP analysis)}. This observation supports our conclusions regarding the importance of a positive feedback by community members, and of link sharing as positive influences on the engagement and cohesiveness of online communities (see Section \ref{subsec:feature_importance}).

\subsection{Performance of the Neural Models}
\label{subsec:performance_of_neural_models}
 Neural models are considered to be the state-of-the-art in many classification and prediction tasks. All the neural architectures we experimented with achieved decent results, significantly outperforming a naive baseline (Table \ref{table:results}). However, the neural models were outperformed by the GBT. We briefly consider the factors that may have affected the performance of the models, though a proper experiment is out of the scope of this paper:
 
\begin{itemize}[leftmargin=0.3cm]
    \item \textit{Dataset size}: Although the size of the dataset ($S^+ \cup S^-$) is considerable (tens of millions of posts submitted by over ten million users), the total number of \emph{instances} is relatively small ($<2500$). 
    Deep learning models perform well on large datasets, with high number of instances that allows a productive back-propagation of error terms to a large number of neurons. 
    \item \textit{Data complexity}:  While the number of instances is relatively small, each instance is a complex unit, composed of thousands of words, threads and users, making feature definition and extraction less trivial.
    \item \textit{Neural architectures}: The neural architectures we designed are aimed to process very different feature types -- word tokens (at the post, thread and community level), meta features, and social features. Optimizing a neural architecture to handle multiple feature types is not a trivial task \cite{sharma2020semeval, wang2020deep}. 
\end{itemize}


\section{Conclusions and Future Work}
\label{sec:closure}
In this work we study how community structure, norms and language can be used to predict the community's engagement in a large scale distributed campaign. Specifically, we predict which Reddit community is likely to take place in the \rp~experiment.
We use Gradient Boosted Trees and complex neural models, experimenting with various representations of community (language, network, meta). We demonstrated that all type of features contribute to the classification, that feature types enhance each other and that meta features are as important as linguistic features. 

Future work takes two trajectories: 
(i) Analysing the rich community dynamics observed \emph{during} the experiment and model the communities' success level, and (ii) Improving the ways complex meta and network features are defined, extracted and introduced into the learning models.

\bibliographystyle{ACM-Reference-Format}
\bibliography{www_bib}

\newpage
\appendix

\section{Algorithmic Approaches}
\label{sec:appendix_alg_approaches}

In this work we use comments data only through meta-features creation and \emph{not} as part of the textual input data for each model. We assume that submissions texts well represent communities' properties and characteristics. We limit to the number of submission texts per community that are used by the model to 10K. This limitation is due to the high complexity of the models when handling thousands of potential long text submissions.\footnote{Each submission may consist of a very high number of sentences and tokens (unlimited number in Reddit).}

As mentioned in Section \ref{sec:characterizing_subreddits}, we use embedding vectors for training the neural models. We considered two embeddings alternatives: an off-the-shelf pretrained model \cite{mikolov2013efficient} and a dedicated GloVe model \cite{pennington2014glove} trained on $DS1$. All reported results were obtained with the latter model which achieved better performance.

In the rest of this section, we elaborate about the classification models we experiment with.

\paragraph{Gradient Boosted Trees (GBT)} We experiment with a various non-sequential classifiers (classifiers that ignore the sequential structure of the data) -- logistic regression, decision tree, SVM, naive Bayes, ada-boost, random-forest, and gradient boosted trees (BGT). However, we report only on the performance of the GBT model, the best performing non-sequential classifier.

GBT is a class of information-theoretical discriminative classifiers. A series of weak learners (decision trees) is constructed, boosting classification accuracy by combining the respective learners, trained on modified and over/under-sampled data \cite{schapire1990strength,friedman2002stochastic}.
GBT classifiers tend to work well on relatively small datasets and with a combination of different and unnormalized feature types. Hence, the GBT classifier seem to be well suited to the task at hand. We limited the number of trees to 100 and the depth of each tree to 3.

\paragraph{Multulayer Perceptron (MLP)}
MLPs are a class of simple feed-forward artificial neural network classifiers, with at least one hidden layer of neurons, successfully learn nonlinear functions \cite{rumelhart1985learning}.

\paragraph{BERT Zero-Shot}
Bidirectional Encoder Representations from Transformers (BERT) \cite{devlin2018bert} is a transformer-based technique that achieves state-of-the-art results in many NLP domains. Most of the BERT models are applicable for short snippets pf text \cite{liu2019roberta, sanh2019distilbert}. Lately, new BERT variants were suggested to handle longer texts \cite{zaheer2020big, beltagy2020longformer}. However, none of them is able to handle the high number of submissions and comments that is generated by each community.
We use BERT representation of sentences in a zero-shot manner \cite{yin2019benchmarking}. We use a pre-trained BERT model to get a vector representation of each submission in the corpus. We use the last layer of the BERT model, of size 768, for this purpose. 
This series of \textit{submission embeddings} is averaged to represent the whole subreddit. The averaged \textit{submission embeddings} is fed to the an \textit{MLP component} along with a vector of the network and the meta features. Its dimension depends on the experimental setting (see  Section \ref{sec:exp_setup}). Finally, a binary \textit{softmax} function is applied in order to predict the class. 

\begin{figure}
    \subfloat[{Parallel LSTM}]{{ \includegraphics[scale=0.39]{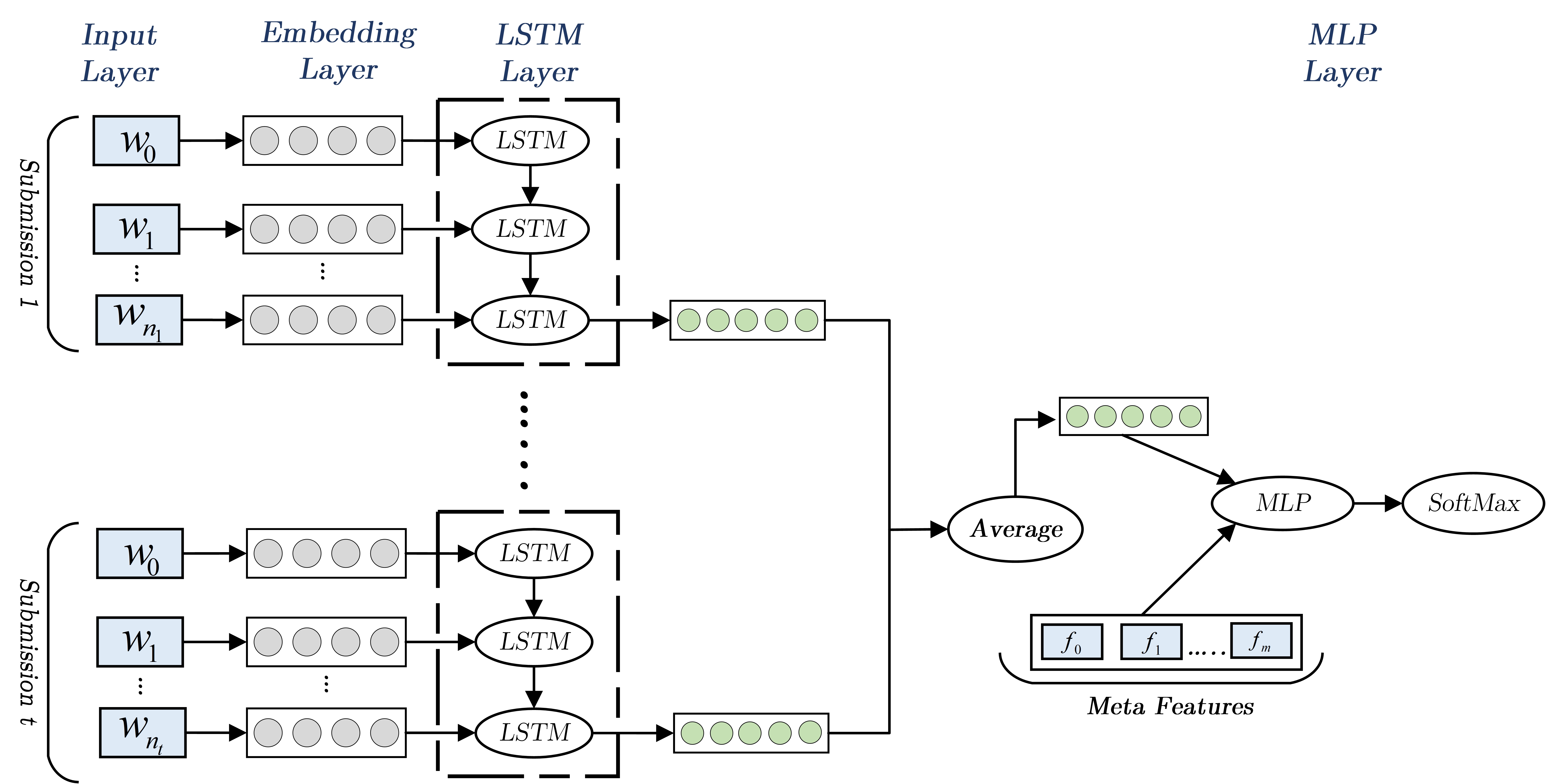}}
    \label{fig:parallel_lstm}}
    \qquad
    \subfloat[CNN Max-Pooling]
    {{\includegraphics[scale=0.345]{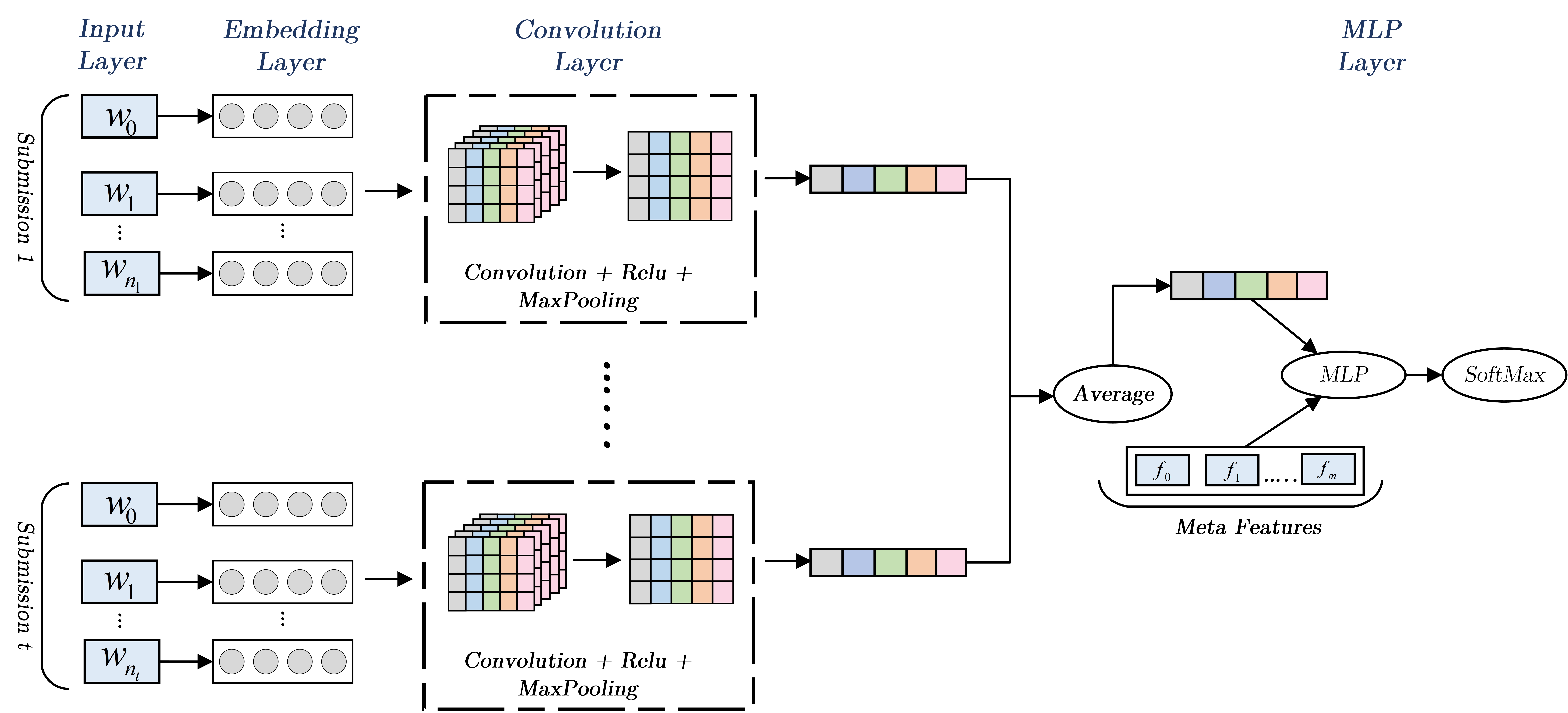}}
    \label{fig:CNN_max_pooling}}
    \caption{\normalsize Parallel LSTM and CNN Max-Pooling architectures. Each input submission is splitted into tokens, and represented as an embedding vector. All Submissions are aggregated into one vector at the end of the LSTM/CNN layer, and then concatenated with the meta features input vector. Dimension of all inputs (number of submissions, length of each submission, and number of meta features) are dynamic. Last layer includes a standard MLP and a softmax part. Blue shaded variables indicate model's input.}
    \Description{Parallel LSTM and CNN Max-Pooling neural modals architecture}
    \label{fig:dl_arch}
\end{figure}

\begin{table*}
\centering
\footnotesize
\caption{Meta and social network features. We calculate each feature per community (subreddit). For the last four features in the table (marked as <feature-name>$^*$), we use  average, maximum, minimum, median, and standard deviation measures.}
{
    \begin{tabular}{l@{\quad}|l@{\quad}|l@{\quad}|l@{\quad}}
      {} & 
      {Feature Name} &
      {Type} &
      {Explanation} \\[2pt]
      \hline\rule{0pt}{12pt}
       \multirow{25}{*}{\vtop{\hbox{\strut Meta}\hbox{\strut Features}}}
        &submission amount & Int & Number of submissions\\[2pt]
        &submission amount normalized   & Float & Number submissions \% number of users\\[2pt]
        &submission average score       & Float & Submission up-vote average score\\[2pt]
        &submission median score        & Float & Submission up-vote median score\\[2pt]
        &comments average score         & Float & Comments up-vote average score\\[2pt]
        &comments median score          & Float & Comments up-vote median score\\[2pt]
        &comments submission ratio      & Float & Number of comments \% number of submissions\\[2pt]
        &deleted removed submission ratio & Float & Number of deleted or removed submission \% number of submissions\\[2pt]
        &distinct comments to submission ratio  & Float &Percentage of submissions which were commented (at least once)\\[2pt]
        &distinct comments to comments ratio  & Float &Number of submissions commented \% number of comments   \\[2pt]
        &users amount                   & Int & Number of users registered (not necessarily wrote anything)  \\[2pt]
        &submission distinct users      & Int & Distinct number of users who wrote a submission (i.e., started a thread)\\[2pt]
        &average submission per user    & Float & Average number of submissions per user (out of those who wrote something)  \\[2pt]
        &median submission per user     & Float & Median number of submissions per user (out of those who wrote something)   \\[2pt]
        &submission to comments users ratio & Float & Distinct number of commented \% distinct number of users who wrote a submission \\[2pt]
        &submission users std           & Float & Users standard deviation, out of the users who wrote a submission  \\[2pt]
        &comments users std             & Float & Users standard deviation, out of the users who wrote a comment     \\[2pt]
        &users deleted normalized       & Float & Number of submission or comments deleted user \% number of total users \\[2pt]
        &submission title length        & Float & Average length of a submission title   \\[2pt]
        &median submission title length & Float & Median length of a submission title\\[2pt]
        &submission selftext length     & Float & Average length of a submission selftext (content of the submission)    \\[2pt]
        &median submission selftext length  & Float & Median length of a submission selftext (content of the submission)   \\[2pt]
        &empty selftext ratio               & Float & Percentage of submissions with an empty selftext   \\[2pt]
        &submissions2comments words used    & Float & Number of comments distinct words \% number of submission distinct words\\[2pt]
        &age(days)                      & Int & Number of days which the community exists (counting back from 31/3/2017)   \\[2pt]
    \hline\rule{0pt}{12pt}
    \multirow{16}{*}{\vtop{\hbox{\strut Social}\hbox{\strut Network}\hbox{\strut Features}}}
        &num of nodes & Int & The overall number of active users)\\[2pt]
        &num of triangles & Int & Number of triads\\[2pt]
        &num of edges & Int &The total number of edges in the graph\\[2pt]
        &is biconnected & Binary & Is the graph biconnected\\[2pt]
        &num of nodes to cut & Int & The min-cut value of the graph's biggest component\\[2pt]
        &density & Float & Number of nodes \% number of edges\\[2pt]
        &num connected components& Int & The number of connected components in the graph\\[2pt]
        &num connected components > 2 & Int & The number of connected components that contain more than two nodes\\[2pt]
        &max group in a connected components & Int & The largest connected component size\\[2pt]
        &num  s.connected components& Int & The number of strongly connected components in the graph\\[2pt]
        &num strongly connected components > 2& Int & The number of strongly connected components that contain more than two nodes\\[2pt]
        &max group in a s.connected components & Int &The largest strongly connected component group size\\[2pt]
        &betweenness$^*$ & Float & Each node's betweenness\\[2pt]
        &centrality$^*$ & Float & Each node's centrality\\[2pt]
        &closeness$^*$ & Float & Each node's closeness value\\[2pt]
        &in degree$^*$ & Int/Float & Each node's in-degree value\\[2pt]
    \hline
    \end{tabular}
}
    \label{table:meta_and_network_feautures_description}
\end{table*}

\paragraph{Parallel LSTM with an MLP Layer}
Long-Short-Term Memory (LSTM) \cite{hochreiter1997long} are special type of recurrent neural network (RNN). The network contains internal loops, allowing an adaptive memory effect. LSTM networks are proven to perform well on many language-related classification tasks as the sequential nature of language and its dependencies are being captured accurately by a series of LSTM cells. Beyond the sequential nature of language, the sequential nature of discussion threads promotes the use of LSTMs.

We design the network to take different types of features and combine them in different layers. An illustration of the neural architecture is presented in Figure \ref{fig:parallel_lstm}. Each input iteration operates on a subreddit. A subreddit is composed of multiple submissions.\footnote{Each submission is the root of a discussion thread, see Section \ref{sec:intro}} The input layer is dynamic, depending on the number of submissions in the current subreddit. Each tokenized submission (i.e., $w_0, w_1...$) is translated to a sequence of \textit{embedding vectors} that are fed into a respective sequence of \textit{LSTM} cells. The LSTM layers yield a series of \textit{submission embeddings} (note -- each submission is a thread in the subreddit) which is averaged to represent the whole subreddit (dark green circles in Figure \ref{fig:parallel_lstm}). The \textit{submission embedding} is fed to the an \textit{MLP component} along with a vector of the network and the meta features. Its dimension depending on the experimental setting (see  Section \ref{sec:exp_setup}). Finally, a binary \textit{softmax} function is applied in order to predict the class.

Note that meta-features are related to a subreddit as a whole, hence the vector of the meta-features is added just before the final MLP component, as illustrated in Figure \ref{fig:parallel_lstm}.

\paragraph{Max pooling CNN with an MLP Layer} Convolutional neural network (CNN) \cite{krizhevsky2012imagenet} are a special type of multilayer perceptron, which apply different convolutions to the input data. Recently, CNNs \ignore{were originally developed to solve computer vision related tasks and lately} proved useful in NLP related problems \cite{ kim2014convolutional, goldberg2017neural}. The architecture we designed is presented in Figure \ref{fig:CNN_max_pooling} and contrasted with the LSTM architecture. The main difference is our usage of a convolutional layer instead of a sequence of LSTM cells. Such convolution step, allows us to capture relations between words sequences (in the embedding form). Empirically, a long width and a short length mask yielded the best results. We use a mask of (100, 2) and a 300 dimensions embedding vector.

 \ignore{Additional approach which was tested, is modeling the \textit{sequences of authors} inside a discussion thread. This means that instead of taking the sequences of tokens and submission in the input layer, we are taking the sequence of authors (i.e., based on time when the submission/comment was written) as input. This semi approach removes the linguistic meaning of the input, and takes into account the structure and content \textbf{of authors} in a community. Such approach did not provide better results compared others and hence will not be presented in Section \ref{sec:res_and_analysis}.
}

\section{Meta and Network Features}
\label{sec:appendix_meta_and_network_features}
As mentioned in Section \ref{sec:characterizing_subreddits}, we use meta features (e.g., comments amount) as well as social network features per community. These network and meta features are commonly used in social network analysis (SNA) research. We preferred \emph{not to} represent the network and meta features as embedding vectors, in order to gain useful insights from the model’s interpretation.

In Table \ref{table:meta_and_network_feautures_description} we provide the full list of 25 meta features followed by the 32 social network features, with a short description of each.
\end{document}